\documentclass[twocolumn,showpacs,superscriptaddress,amsmath,amssymb,aps,prl]{revtex4-1}
\usepackage{graphicx}
\usepackage{dcolumn}
\usepackage{amsmath,bm}
\usepackage{epstopdf}
\usepackage[mathlines]{lineno}
\usepackage{color}
\usepackage[colorlinks=true,linkcolor=blue,citecolor=blue]{hyperref}

\begin{document}
\title{Controlling a Nanowire Spin-Orbit Qubit via Electric-Dipole Spin Resonance}
\author{Rui Li}
\affiliation{Beijing Computational Science Research Center, Beijing
100084, China}
\affiliation{Department of Physics and State Key Laboratory of
Surface Physics, Fudan University, Shanghai 200433, China}

\author{J. Q. You}
\affiliation{Beijing Computational Science Research Center, Beijing
100084, China}
\affiliation{Department of Physics and State Key Laboratory of
Surface Physics, Fudan University, Shanghai 200433, China}
\affiliation{Center for Emergent Matter Science, RIKEN, Saitama
351-0198, Japan}

\author{C. P. Sun}
\affiliation{Beijing Computational Science Research Center, Beijing
100084, China}
\affiliation{Center for Emergent Matter Science, RIKEN, Saitama
351-0198, Japan}

\author{Franco Nori}
\affiliation{Center for Emergent Matter Science, RIKEN, Saitama 351-0198,
Japan}
\affiliation{Physics Department, The University of Michigan,
Ann Arbor, Michigan 48109-1040, USA}

\begin{abstract}
A semiconductor nanowire quantum dot with strong spin-orbit coupling (SOC) can be used to achieve a spin-orbit qubit. In contrast to a spin qubit, the spin-orbit qubit can respond to an external ac lectric field, an effect called electric-dipole spin resonance. Here we develop a theory that can apply in the strong SOC regime. We find that there is an optimal SOC strength $\eta_{\rm opt}=\sqrt{2}/2$, where the Rabi frequency induced by the ac electric field becomes maximal. Also, we show that both the level spacing and the Rabi frequency of the spin-orbit qubit have periodic responses to the direction of the external static magnetic field. These responses can be used to determine the SOC in the nanowire.
\end{abstract}
\date{\today}
\maketitle

{\it Introduction}.---How to achieve a simple and efficient way to manipulate a qubit is of basic importance in quantum information processing (see, e.g., \cite{Buluta,Xiang}). For the conventional spin qubit~\cite{Hanson}, its manipulation can be accomplished by using the electron spin resonance technique~\cite{Awschalom,Koppens,Engel,Leon,Sanchez}. The spin-orbit qubit~\cite{Nadj1}, unlike the conventional spin qubit, contains both the orbital and the spin degrees of freedom of an electron, owing to the spin-orbit coupling (SOC)~\cite{Winkler}. The spin-orbit qubit has an additional advantage of being manipulable via an external ac electric field, an interesting phenomenon called the electric-dipole spin resonance (EDSR)~\cite{Nowack,Pioro,Tokura,Laird,Bednarek,Nadj2}. With respect to generating a local ac magnetic field for manipulating a spin qubit, it is much easier to produce a local ac electric field with current experimental techniques.

The prerequisite for realizing a spin-orbit qubit in a semiconductor quantum-dot structure is the availability of SOC in the material~\cite{Nadj1}. There are two different types of SOC in a semiconductor material, i.e., the Rashba SOC due to structural inversion asymmetry~\cite{bychkov2} and the Dresselhaus SOC due to the bulk inversion asymmetry~\cite{Dresselhaus}. Usually, both types of SOC coexist in a material~\cite{Voskoboynikov}, but which one plays a major part depends on the properties of the material.

Semiconductor quantum wires with strong SOC, e.g., InSb nanowires~\cite{Nadj1,Nadj2}, are of current interest. These have been suggested as a potential platform for demonstrating Majorana quasiparticles~\cite{Lutchyn,Oreg}, and these can also be used to produce a quantum dot for achieving a spin-orbit qubit~\cite{Nadj1}. The coherent electric manipulation and the spectroscopy of a nanowire spin-orbit qubit were investigated~\cite{Nadj2}, and a strong Rabi frequency of 100 MHz was also reported recently~\cite{van}. Interestingly, the frequency of the driving ac electric field depends on the direction of the applied static magnetic field~\cite{Nadj2}. As our study shows, this dependence is actually a signature of the strong SOC in the nanowire.

In this Letter, we provide an explicit theoretical explanation for the EDSR effect in a nanowire quantum dot with strong SOC. In comparison with previous theories, where the SOC was regarded as a perturbation~\cite{Rashba,Golovach,Rashba3,Hu}, we consider a strong SOC. Instead, we treat the external static magnetic field as a perturbation. It is estimated that our theory can be valid when using a magnetic field as strong as 0.1 T. This field is much stronger than the magnetic field usually used in experiments on quantum devices. With our theory applicable in the strong SOC regime, it reveals that the Rabi frequency induced by an external ac electric field has a maximum value at an optimal SOC strength, instead of the Rabi frequency that is linearly proportional to the SOC strength. As our theory shows, this linear dependence is only valid in the weak SOC regime. Also, our theory shows that the SOC can be probed by monitoring both the spectrum and the EDSR responses of the spin-orbit qubit to the direction of the external static magnetic field. Therefore, it can provide a useful method to determine both Rashba and Dresselhaus SOCs in the nanowire.

\begin{figure}
\includegraphics{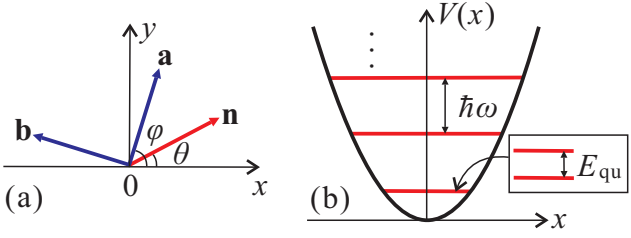}
\caption{\label{Fig_angle}(color online) (a)~Schematic diagram of the unit vectors $\textbf{n}$, $\textbf{a}$, and $\textbf{b}$, where $\textbf{n}=(\cos\theta,\sin\theta)$ represents the direction of an external static magnetic field $\textbf{B}$, $\textbf{a}=(\cos\varphi,\sin\varphi)$, with $\varphi=\arctan(\alpha_{\rm R}/\alpha_{\rm D})\in[0,\frac{\pi}{2}]$ characterizing the relative strength between the Rashba and Dresselhaus SOCs, and $\textbf{b}=(-\sin\varphi,\cos\varphi)$ is a unit vector perpendicular to $\textbf{a}$. (b)~Energy spectrum of a nanowire quantum dot modeled by $H_{0}$ in Eq.~(\ref{Eq Two}), where each energy level is twofold degenerate. This degeneracy can be removed by applying a static magnetic field to the nanowire quantum dot.  Here the lowest two levels with splitting $E_{\rm qu}$ are used to encode a spin-orbit qubit.}
\end{figure}

{\it Spin-orbit qubit based on a nanowire quantum dot}.---We consider a gated nanowire quantum dot with strong SOC, where an electron is confined in an 1D harmonic well and subject to a static magnetic field~\cite{Pershin,Nowak}. The Hamiltonian reads
\begin{equation}
H=\frac{p^{2}}{2m_{e}}+\frac{1}{2}m_{e}\omega^{2}x^{2}+\alpha_{\rm R}\sigma^{y}p+\alpha_{\rm D}\sigma^{x}p
+\frac{g_{e}\mu_{\rm B}B}{2}\sigma^{n},\label{Eq_One_D_Hamiltonian}
\end{equation}
where $m_{e}$ is the effective electron mass, $p=-i\hbar\partial/\partial\,x$, $\alpha_{\rm R(D)}$ is the Rashba (Dresselhaus) SOC strength, $\mu_{\rm B}$ is Bohr magneton, and $\sigma^{n}\equiv\textbf{n}\cdot\boldsymbol{\sigma}=\sigma^{x}\cos\theta+\sigma^{y}\sin\theta$, with $\textbf{n}=(\cos\theta,\sin\theta)$ representing the direction of the external static magnetic field [see Fig.~\ref{Fig_angle}(a)].

Even though the Hamiltonian (\ref{Eq_One_D_Hamiltonian}) looks simple, it is very difficult to analytically calculate its energy spectrum by directly solving the Schr{\"o}dinger equation~\cite{Bulgakov,Tsitsishvili,Rashba2}. Therefore, in order to have a good understanding of the energy spectrum and the corresponding eigenstates of $H$, one has to rely on a perturbative method. Here we introduce two new parameters $\alpha$ and $\varphi$: $\alpha=\sqrt{\alpha^{2}_{\rm R}+\alpha^{2}_{\rm D}}$, and $\varphi=\arctan(\alpha_{\rm R}/\alpha_{\rm D})$, where $\varphi\in[0,\frac{\pi}{2}]$. With these new parameters, the Hamiltonian (\ref{Eq_One_D_Hamiltonian}) can be rewritten as
\begin{eqnarray}
H&=&H_{0}+H_{1},\nonumber\\
H_{0}&=&\frac{p^{2}}{2m_{e}}+\frac{1}{2}m_{e}\omega^{2}x^{2}+\alpha\sigma^{a}p,\nonumber\\
H_{1}&=&\frac{g_{e}\mu_{\rm B}B}{2}[\cos(\theta-\varphi)\sigma^{a}+\sin(\theta-\varphi)\sigma^{b}],
\label{Eq Two}
\end{eqnarray}
with $\sigma^{c}\equiv\textbf{c}\cdot\boldsymbol{\sigma}$ ($\textbf{c}=\textbf{a},\textbf{b},\textbf{n}$).
Here the unit vectors $\textbf{a}=(\cos\varphi,\sin\varphi)$, $\textbf{b}=(-\sin\varphi,\cos\varphi)$, and
$\textbf{n}=\cos(\theta-\varphi)\textbf{a}+\sin(\theta-\varphi)\textbf{b}$ are schematically illustrated in Fig.~\ref{Fig_angle}(a). In previous theories~\cite{Rashba,Golovach,Rashba3,Hu}, the SOC term  was often considered as a perturbation, but this applies only for a weak SOC, i.e., $\eta\equiv\sqrt{m_e/(\hbar\omega)}\alpha\ll 1$. Below we consider the case which is valid even in the strong SOC (i.e., large $\eta$) regime. Note that the Zeeman splitting $g_{e}\mu_{\rm B}B$ ($\sim1$ $\mu$eV) is usually much less than the orbit splitting $\hbar\omega$ ($\sim~1$--$10$ meV). For instance, in an InSb nanowire quantum dot, $\hbar\omega\approx7.5$ meV and $g_{e}\approx40$~\cite{Nadj2}, so the external magnetic field can be as strong as $B\sim0.1$ T for $\xi\equiv g_{e}\mu_{\rm B}B/(\hbar\omega)\sim0.03$. This field is much stronger than the magnetic field usually used in  experiments~\cite{Nadj1,Nadj2,van}, so $H_{1}$ can be treated as a perturbation.

To encode a spin-orbit qubit, we only need to focus on the lowest two energy levels of $H$. Here we calculate the energy-level spacing and the Hilbert-space structure of the spin-orbit qubit by using the perturbative method for degenerate states~\cite{Landau}, where all derived results are accurate up to first order in $\xi$.

The Hamiltonian $H_{0}$ in Eq.~(\ref{Eq Two}) can be diagonalized using a unitary transformation~\cite{Levitov}, i.e., $e^{i(m_e\alpha/\hbar)x\sigma^{a}}H_{0}e^{-i(m_e\alpha/\hbar)x\sigma^{a}}=\frac{p^{2}}{2m_{e}}
+\frac{1}{2}m_{e}\omega^{2}x^{2}-m_e\alpha^2/2$. Let $\psi_{n}(x)$ be the eigenstates of a harmonic oscillator corresponding to the eigenvalues $(n+\frac{1}{2})\hbar\omega$, where $n=0,1,2,\dots$. Then, the eigenvalues of $H_{0}$ are $\varepsilon_{n}=(n+\frac{1}{2})\hbar\omega-\frac{1}{2}m_e\alpha^2$, and the corresponding eigenstates are given by
\begin{equation}
  \left\{\begin{array}{l}
        |\Psi_{n\uparrow}\rangle=e^{-i(m_e\alpha/\hbar)x}\psi_{n}(x)|\!\uparrow_{a}\rangle,\\
        |\Psi_{n\downarrow}\rangle=e^{i(m_e\alpha/\hbar)x}\psi_{n}(x)|\!\downarrow_{a}\rangle,
      \end{array}
      \right.
\end{equation}
with $|\!\!\!\uparrow_{a}\rangle=(\sqrt{2}/2)(e^{-i\varphi/2}, e^{i\varphi/2})^{T}$, and $|\!\!\!\downarrow_{a}\rangle=(\sqrt{2}/2)(e^{-i\varphi/2},-e^{i\varphi/2})^{T}$, where $T$ denotes the transpose of a matrix, being two eigenstates of $\sigma^{a}$: $\sigma^{a}|\!\uparrow_{a}\rangle=|\!\uparrow_{a}\rangle$, and $\sigma^{a}|\!\downarrow_{a}\rangle=-|\!\downarrow_{a}\rangle$. The energy spectrum of $H_{0}$ is similar to the energy spectrum of a harmonic oscillator, except that each level $\varepsilon_{n}$ is twofold degenerate,
with the corresponding degenerate eigenstates given by $|\Psi_{n\uparrow}\rangle$ and $|\Psi_{n\downarrow}\rangle$.

Our interest focuses on the $n=0$ Hilbert subspace.
Usually, the two degenerate states $|\Psi_{0\uparrow}\rangle$ and $|\Psi_{0\downarrow}\rangle$ will recombine in the zeroth-order wave functions, so we  calculate $H_{1}$ in the Hilbert subspace spanned by $|\Psi_{0\uparrow}\rangle$ and $|\Psi_{0\downarrow}\rangle$:
\begin{equation}
H_{1}=\frac{g_{e}\mu_{\rm B}B}{2}\left(
                  \begin{array}{cc}
\cos(\theta-\varphi) & ie^{-\eta^{2}}\sin(\theta-\varphi) \\
                    -ie^{-\eta^{2}}\sin(\theta-\varphi) & -\cos(\theta-\varphi) \\
                  \end{array}
                \right).
\end{equation}
Here we have used the formulas $\sigma^{b}|\!\uparrow_{a}\rangle=-i|\!\downarrow_{a}\rangle$ and $\sigma^{b}|\!\downarrow_{a}\rangle=i|\!\uparrow_{a}\rangle$ in deriving the above matrix. Diagonalizing this matrix, we obtain the eigenvalues and the corresponding eigenfunctions
\begin{equation}
\begin{array}{ll}
  \varepsilon^{\pm}_{0p}=\pm\frac{g_{e}\mu_{\rm B}B}{2}f, & ~~|\Psi^{\pm}_{0}\rangle=c^{\pm}_{0}|\Psi_{0\uparrow}\rangle+d^{\pm}_{0}|\Psi_{0\downarrow}\rangle,
\end{array}
\end{equation}
where
\begin{eqnarray}
c^{\pm}_{0}&=&\frac{\cos(\theta-\varphi)\pm\,f}{\sqrt{2[f^{2}\pm\,f\cos(\theta-\varphi)]}},\nonumber\\
d^{\pm}_{0}&=&\frac{-ie^{-\eta^{2}}\sin(\theta-\varphi)}{\sqrt{2[f^{2}\pm\,f\cos(\theta-\varphi)]}},
\end{eqnarray}
with $f\equiv\,f(\eta,\theta-\varphi)=[\cos^{2}(\theta-\varphi)+e^{-2\eta^{2}}\sin^{2}(\theta-\varphi)]^{1/2}$.
The wave functions $|\Psi^{\pm}_{0}\rangle$ are actually the recombined zeroth-order wave functions. The first-order wave functions can be calculated using the perturbative formula~\cite{Landau}: $|\Psi^{\pm}_{0p}\rangle=|\Psi^{\pm}_{0}\rangle+\sum^{\infty}_{n=1,\sigma=\uparrow,\downarrow}
\frac{\langle\Psi_{n\sigma}|H_{1}|\Psi^{\pm}_{0}\rangle}{\varepsilon_{0}
-\varepsilon_{n}}|\Psi_{n\sigma}\rangle$. Therefore, we obtain
\begin{eqnarray}
|\Psi^{\pm}_{0p}\rangle&=&c^{\pm}_{0}|\Psi_{0\uparrow}\rangle+d^{\pm}_{0}|\Psi_{0\downarrow}\rangle
+i\frac{\xi}{2}e^{-\eta^{2}}\sin(\theta-\varphi)\nonumber\\
&&\times\sum^{\infty}_{n=1}\frac{\left(\sqrt{2}i\eta\right)^{n}}{n\sqrt{n!}}\left[(-1)^{n}c^{\pm}_{0}
|\Psi_{n\downarrow}\rangle-d^{\pm}_{0}|\Psi_{n\uparrow}\rangle\right].~~~
\label{Eq_wavefunction}
\end{eqnarray}
Here we have obtained the two lowest energy levels and the corresponding wave functions $|\Psi^{\pm}_{0p}\rangle$ of $H$ by using degenerate perturbation theory, which are accurate up to first order in $\xi$.
The two states $|\Psi^{+}_{0p}\rangle$ and $|\Psi^{-}_{0p}\rangle$ can be used to encode the spin-orbit qubit which has
the level spacing [see Fig.~\ref{Fig_angle}(b)]:
\begin{equation}
E_{\rm qu}=g_{e}\mu_{\rm B}B\sqrt{\cos^{2}(\theta-\varphi)+e^{-2\eta^{2}}\sin^{2}(\theta-\varphi)}.
\label{level spacing}
\end{equation}
Clearly, it can be seen from Eq.~(\ref{Eq_wavefunction}) that the spin-orbit qubit is different from the conventional spin qubit (which only contains the $n=0$ orbit state) because the spin-orbit qubit combines many orbital states ($n=0,1,\dots,\infty$) with the spin state. This orbital feature of the spin-orbit qubit leads to an interesting phenomenon called EDSR, which can be used to manipulate the spin-orbit qubit via an external a.c.~electric field. The spin-orbit qubit can be regarded as a hybrid qubit which contains both the orbital and the spin degrees of freedom of an electron in a quantum dot. Therefore, the spin-orbit qubit can respond to both magnetic and electric fields.

Note that the results obtained so far are based on a 1D harmonic quantm well for a single nanowire quantum dot. Usually, the real well may include some anharmonicity and the nanowire is quasi-1D. In this case, the problem cannot be analytically solved, but the underlying physics should be the same because the feature of well-separated discrete orbital levels remains unchanged in the nanowire quantum dot.

{\it EDSR and its response to the magnetic-field direction}.---In EDSR, the SOC plays a key role~\cite{Tokura,Laird,Bednarek,Nadj2,Flindt,Khomitsky,Ban}, such that an electron spin can respond to an ac electric field.
In previous studies, the SOC was treated as a perturbation, and the EDSR effect has been investigated for a quantum-well structure~\cite{Rashba} and a 2D GaAs quantum dot~\cite{Golovach,Rashba3}. Those results show that the Rabi frequency is linearly proportional to the SOC strength $\eta$~\cite{Golovach}.
Below we will show that this linear dependence of the Rabi frequency on $\eta$ is due to the weak SOC.

When we apply an external ac electric field to the nanowire quantum dot, the total Hamiltonian becomes
\begin{eqnarray}
H_{\rm tot}&=&\frac{p^{2}}{2m_{e}}+\frac{1}{2}m_{e}\omega^{2}x^{2}+\alpha_{\rm R}\sigma^{y}p
+\alpha_{\rm D}\sigma^{x}p+\frac{g_{e}\mu_{\rm B}B}{2}\sigma^{n}\nonumber\\
&&+eEx\cos(2\pi\nu t),\label{Eq_One_D_Hamiltonian2}
\end{eqnarray}
where $\nu$ is the frequency of the ac electric field. Based on the results derived above, when we focus on the Hilbert subspace of the spin-orbit qubit spanned by $|\Psi^{+}_{0p}\rangle$ and $|\Psi^{-}_{0p}\rangle$, the Hamiltonian is reduced to a spin-orbit qubit interacting with an ac electric field: $H_{\mathrm{tot}}=\frac{1}{2}E_{\rm qu}\tau^{Z}+eEx\cos(2\pi\nu t)$,
where $\tau^{Z}=|\Psi^{+}_{0p}\rangle\langle\Psi^{+}_{0p}|-|\Psi^{-}_{0p}\rangle\langle\Psi^{-}_{0p}|$. In this Hilbert subspace of the spin-orbit qubit, $x$ has the elements
\begin{eqnarray}
&&\langle\Psi^{+}_{0p}|x|\Psi^{+}_{0p}\rangle=0+\mathcal{O}(\xi^{2}),
\quad\langle\Psi^{-}_{0p}|x|\Psi^{-}_{0p}\rangle=0+\mathcal{O}(\xi^{2}),
\nonumber\\
&&\langle\Psi^{+}_{0p}|x|\Psi^{-}_{0p}\rangle=-ix_{0}\xi\eta\,e^{-\eta^{2}}\sin(\theta-\varphi)+\mathcal{O}(\xi^{2}),
\end{eqnarray}
where $x_0=\sqrt{\hbar/(m_e\omega)}$. Thus, we conclude that $H_{\rm tot}$ can be reduced to the following EDSR Hamiltonian:
\begin{equation}
H_{\rm tot}=(1/2)E_{\rm qu}\tau^{Z}+h\Omega_{\rm R}\tau^{Y}\cos(2\pi\nu t),
\end{equation}
where
\begin{equation}
\Omega_{\rm R}=\left(\frac{eEx_0}{h}\right)\xi\eta\,e^{-\eta^{2}}|\sin(\theta-\varphi)|,
\label{Rabi}
\end{equation}
is the Rabi frequency, with $h$ being the Planck constant, and $\tau^{Y}=i(|\Psi^{-}_{0p}\rangle\langle\Psi^{+}_{0p}|-|\Psi^{+}_{0p}\rangle\langle\Psi^{-}_{0p}|)$. When $\eta$ is treated as a perturbation for weak SOC ($\eta\ll1$), we can expand $\Omega_{\rm R}$ as $\Omega_{\rm R}=(eEx_{0}/h)\xi\eta|\sin(\theta-\varphi)|+\mathcal{O}(\eta^{2})$, so we recover the previous result~\cite{Golovach}. In fact, this EDSR Hamiltonian is the Rabi-oscillation Hamiltonian in quantum optics~\cite{Scully}. Therefore, the physics of EDSR is transparent: when the driving frequency is in resonance to the level spacing of the spin-orbit qubit ($h\nu=E_{\rm qu}$), the spin-orbit qubit can be flipped from one basis state to another via the Rabi oscillation. Note that the Rabi frequency $\Omega_{\rm R}$ is obtained by only considering the lowest two levels of the quantum dot. As we show in \cite{supplementary}, when the resonant condition is satisfied ($h\nu=E_{\rm qu}$) and the driving electric field is weak ($E\ll\hbar\omega/\sqrt{2}ex_{0}$), the influence of other energy levels is negligible.

\begin{figure}
\includegraphics{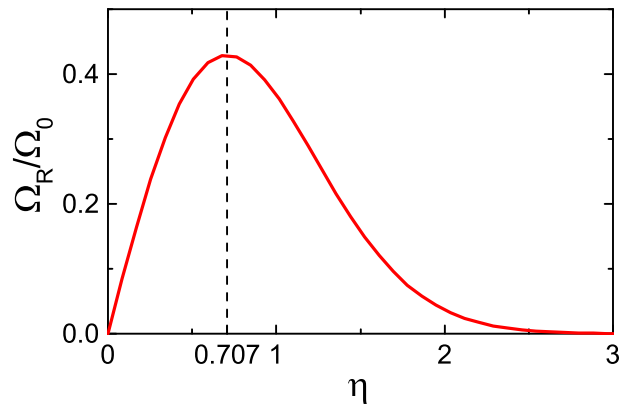}
\caption{\label{Fig_Optimal_SOC}(color online) Rabi frequency (in units of $\Omega_{0}$) versus the SOC strength $\eta$, where $\Omega_{0}=(eEx_{0}/h)\xi|\sin(\theta-\varphi)|$.  There is an optimal SOC strength $\eta_{\rm opt}=\sqrt{2}/2\approx 0.707$, where the Rabi frequency becomes maximal. After this optimal point $\eta_{\rm opt}$, increasing SOC reduces the Rabi frequency.}
\end{figure}

In experiments, the EDSR was probed using a double quantum dot, where the double quantum dot was initially tuned in the Pauli spin blockade regime~\cite{Nadj1,Nowack,Nadj2,Laird}. If and only if an electron spin in either dot is flipped via EDSR, the current can flow through the dot. Here we emphasize that for a nanowire quantum dot with strong SOC, the electron spin in the dot becomes a pseudospin (a spin-orbit qubit). Because the current through the dot depends on the states of the spin-orbit qubit, the flipping of this pseudospin can also be monitored by either the current through the dot~\cite{Nadj1,Nowack,Nadj2} or a nearby quantum point contact~\cite{Laird}.

Note that the SOC strength $\eta$ is not treated as a perturbation in our theory, so our results apply in the strong SOC regime. First, the level spacing $E_{\rm qu}$ given in Eq.~(\ref{level spacing}) depends on the direction $\theta$ of $\textbf{B}$. If $\eta$ is treated as a perturbation as for a weak SOC, the level spacing is just the Zeeman splitting $g_{e}\mu_{\rm B}B$. Thus, the dependence of $E_{\rm qu}$ on the magnetic-field direction is a signature of strong SOC in the nanowire. This directional dependence was demonstrated in a recent experiment on an InSb nanowire quantum dot~\cite{Nadj2}. Second, the Rabi frequency $\Omega_{\rm R}$ given in Eq.~(\ref{Rabi}) is not linearly proportional to the SOC strength $\eta$. Instead, there is an optimal SOC strength $\eta_{\rm opt}=\sqrt{2}/2$, where the Rabi frequency reaches its maximum value (see Fig.~\ref{Fig_Optimal_SOC}).
Our results imply that, in order to achieve the strongest Rabi frequency, it is not necessary to find materials with extremely strong SOC, but to find a material with an optimal SOC strength $\eta_{\rm opt}$. This optimal material gives the smallest manipulation time $\Omega^{-1}_{\rm R}$ for the state flipping of a spin-orbit qubit.

\begin{figure}
\includegraphics[width=3.2in]{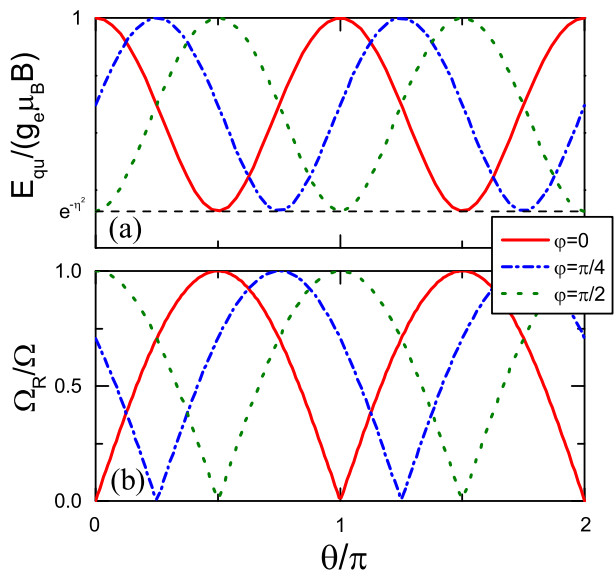}
\caption{\label{Fig_response}(color online) (a)~Periodic response of the level spacing $E_{\rm qu}$ (in units of $g_e\mu_{\rm B}B$) of the spin-orbit qubit on the direction $\theta$ of the external static magnetic field $\textbf{B}$. (b)~Periodic response of the Rabi frequency $\Omega_{\rm R}$ (in units of $\Omega$) on the direction $\theta$ of $\textbf{B}$, where $\Omega=(eEx_{0}/h)\xi\eta\,e^{-\eta^{2}}$. The parameter $\varphi=\arctan(\alpha_{\rm R}/\alpha_{\rm D})$ characterizes the relative strength between the Rashba and Dresselhaus SOCs in the nanowire. For example, $\varphi=0$ corresponds to a nanowire with pure Dresselhaus SOC ($\alpha_{\rm R}=0$), $\varphi=\pi/2$ corresponds to a nanowire with pure Rashba SOC ($\alpha_{\rm D}=0$), and $\varphi=\pi/4$ corresponds to a nanowire with equal Rashba and Dresselhaus SOCs ($\alpha_{\rm R}=\alpha_{\rm D}$).}
\end{figure}

 The determination of the SOC is an important goal~\cite{Fasth}. Because both, level spacing and Rabi frequency of the spin-orbit qubit, depend on the direction of ${\bf B}$, we can use these responses to determine both the Rashba and  Dresselhaus SOC strengths, $\alpha_{\rm R}$ and $\alpha_{\rm D}$, in the nanowire. Figure~\ref{Fig_response} shows the dependence of both the level splitting and the Rabi frequency on the magnetic-field direction for different values of $\varphi=\arctan(\alpha_{\rm R}/\alpha_{\rm D})$. Because the level splitting $E_{\rm qu}$ oscillates between $e^{-\eta^{2}}$ and 1 [see Fig.~\ref{Fig_response}(a)], we can determine the SOC strength $\eta$ from the minimal amplitude $e^{-\eta^{2}}$. Moreover, by monitoring how the level splitting $E_{\rm qu}$ and the Rabi frequency $\Omega_{\rm R}$ vary with the direction $\theta$ of $\textbf{B}$, we can determine the parameter $\varphi$. For example, the level spacing $E_{\rm qu}$ in Eq.~(\ref{level spacing}) reaches its maximum values at $\theta_{\rm max}=\pm\,l\pi+\varphi$ ($l=0,1,2\dots$) [see Fig.~\ref{Fig_response}(a)], and the Rabi frequency $\Omega_{\rm R}$ in Eq.~(\ref{Rabi}) reaches its maximum values at $\theta_{\rm max}=\pm(2l+1)\pi/2+\varphi$ ($l=0,1,2,\dots$) [see Fig.~\ref{Fig_response}(b)]. Thus, $\varphi$ can be determined from the values of $\theta_{\rm max}$. To obtain $\alpha_{\rm D}=\alpha\cos\varphi$, and $\alpha_{\rm R}=\alpha\sin\varphi$, where $\alpha=\eta\sqrt{\hbar\omega/m_{e}}$, we should know the orbit-level spacing $\hbar\omega$. Actually, $\hbar\omega$  is controlled by the gate voltages on the static electric gates which are used to form the trap potential $\frac{1}{2}m_{e}\omega^{2}x^{2}$. We take a recent experiment as an example. By comparing Fig.~\ref{Fig_response}(a) with Fig.~2(c) in \cite{Nadj2}, we obtain $e^{-\eta^{2}}\approx15/20$ and $\theta_{\rm max}\approx0.22\pi$ at $l=0$ from the experimental data. Thus, we have $\eta\approx0.54$ and $\varphi=\theta_{\rm max}\approx0.22\pi$ ($40^{\circ}$) for the nanowire material. Also, this value of $\eta\approx0.54$ reveals that the nanowire material used in \cite{Nadj2} has a strong SOC.

{\it Discussions and conclusions}.---The EDSR that we considered is only induced by the SOC in the quantum dot. Actually, the hyperfine interaction between the electron spin and the lattice nuclear spins in the dot can also induce interesting phenomena such as hyperfine-mediated EDSR~\cite{Laird,Rashba3,Shafiei} and the spin-resonance locking~\cite{Vink}. In some materials, the electron $g_{e}$ factor may have strong anisotropy. When the static magnetic field rotates, this strong anisotropy might also give rise to appreciable directional oscillations of the level spacing $E_{\rm qu}$. However, it does not yield directional oscillations to the Rabi frequency because $g_{e}$ is not included in $\Omega_{\rm R}$. Thus, one can use the Rabi frequency to demonstrate the directional oscillations induced by the SOC. As to directly showing the SOC-induced directional oscillations in $E_{\rm qu}$, one can use materials with a weak anisotropy in the $g_e$ factor.

In conclusion, we have theoretically investigated the EDSR effect in a semiconductor nanowire quantum dot with  strong SOC. In contrast to the previous theories developed in the weak-SOC regime, our results demonstrate that there is an optimal SOC strength $\eta_{\rm opt}=\sqrt{2}/2$ where the Rabi frequency induced by the external ac electric field is maximal. Also, we show that both the level spacing and the Rabi frequency of the spin-orbit qubit have periodic responses to the direction of the external static magnetic field. These responses can be used to probe the SOC in the nanowire by determining both the Rashba and the Dresselhaus SOC strengths in the material.

R.L. and J.Q.Y. are partially supported by the National Basic Research Program of China Grant No. 2009CB929302 and the National Natural Science Foundation of China Grant No. 91121015.
F.N. is partially supported
by the ARO, RIKEN iTHES Project,
MURI Center for Dynamic Magneto-Optics,
JSPS-RFBR Contract No.~12-02-92100,
Grant-in-Aid for Scientific Research (S),
MEXT Kakenhi on Quantum Cybernetics
and the JSPS via its FIRST program.

\begin{widetext}
\section{Supplementary Material for:\\Controlling a nanowire spin-orbit qubit via electric-dipole spin resonance}

In this supplementary material, we study the influence of other energy levels (besides the lowest two levels for a spin-orbit qubit) in a nanowire quantum dot on the electric-dipole spin resonance (EDSR). We show that when the resonant condition is satisfied, i.e., the frequency of the external driving electric field is in resonance with the level spacing of the spin-orbit qubit ($h\nu=E_{\rm qu}$), and when the driving electric field is weak, i.e., $E\ll\hbar\omega/\sqrt{2}ex_{0}$, the effect of other levels on the spin-orbit qubit is negligibly small and the nanowire quantum dot can indeed be used as a two-level quantum system (spin-orbit qubit).

We study the energy spectrum and the corresponding quantum states of the nanowire quantum dot using degenerate perturbation theory. Note that the two states $|\Psi_{n\uparrow}\rangle$ and $|\Psi_{n\downarrow}\rangle$ given in Eq.~(3) of the main text are degenerate. In the Hilbert subspace spanned by these two degenerate states, the perturbation Hamiltonian $H_{1}$ in Eq.~(2) of the main text can be expressed as
\begin{equation}
H_{1}=\frac{g_{e}\mu_{\rm B}B}{2}\left(
  \begin{array}{ll}
\cos(\theta-\varphi) & iL_{n}(2\eta^{2})e^{-\eta^{2}}\sin(\theta-\varphi) \\
    -iL_{n}(2\eta^{2})e^{-\eta^{2}}\sin(\theta-\varphi) & -\cos(\theta-\varphi) \\
  \end{array}
\right),\label{Eq_generalH}
\end{equation}
where the $L_{n}(x)$ are Laguerre polynomials~\cite{Ferraro_s}
\begin{equation}
L_{n}(x)=\sum^{n}_{k=0}\frac{n!}{k!(n-k)!}\frac{(-1)^{k}}{k!}x^{k}.
\end{equation}
For $n=0,1,2,\dots$, $L_0(x)=1$, $L_1(x)=1-x$, $L_2(x)=\frac{1}{2}(x^{2}-4x+2)$, $\dots$.
Diagonalizing Eq.~(\ref{Eq_generalH}), we obtain the eigenvalues, with respect to $(n+\frac{1}{2})\hbar\omega-\frac{1}{2}m_e\alpha^2$, and the corresponding eigenfunctions:
\begin{equation}
\varepsilon^{\pm}_{np}=\pm\frac{g_{e}\mu_{\rm B}B}{2}f_{n},\quad\,|\Psi^{\pm}_{n}\rangle=c^{\pm}_{n}|\Psi_{n\uparrow}\rangle+d^{\pm}_{n}|\Psi_{n\downarrow}\rangle,
\end{equation}
where
\begin{equation}
c^{\pm}_{n}=\frac{\cos(\theta-\varphi)\pm\,f_{n}}{\sqrt{2[f^{2}_{n}\pm\,f_{n}\cos(\theta-\varphi)]}},\quad\,d^{\pm}_{n}=
\frac{-iL_{n}(2\eta^{2})e^{-\eta^{2}}\sin(\theta-\varphi)}{\sqrt{2[f^{2}_{n}\pm\,f_{n}\cos(\theta-\varphi)]}},
\end{equation}
with
\begin{equation}
f_{n}\equiv\,f_{n}(\eta,\theta-\varphi)=\sqrt{\cos^{2}(\theta-\varphi)+L^{2}_{n}(2\eta^{2})e^{-2\eta^{2}}
\sin^{2}(\theta-\varphi)}.
\end{equation}
We note that $|\Psi^{\pm}_{n}\rangle$, $n=0,1,2,\dots$, are actually the zeroth-order wave functions. The first-order wave functions can be calculated using the perturbative formula~\cite{Landau_s}:
\begin{equation}
|\Psi^{\pm}_{np}\rangle=|\Psi^{\pm}_{n}\rangle+\sum^{\infty}_{m\neq\,n,\sigma
=\uparrow,\downarrow}\frac{\langle\Psi_{m\sigma}|H_{1}|\Psi^{\pm}_{n}\rangle}
{\varepsilon_{n}-\varepsilon_{m}}|\Psi_{m\sigma}\rangle.
\end{equation}

For the $n=0$ orbital, when the spin-orbit coupling is included, the two energy levels (with respect to $\frac{1}{2}\hbar\omega-\frac{1}{2}m_e\alpha^2$) and the corresponding wave functions are given by Eqs.~(4)-(7) of the main text. For the $n=1$ orbital, when the spin-orbit coupling is included, the two energy levels are given, with respect to $\frac{3}{2}\hbar\omega-\frac{1}{2}m_e\alpha^2$, by
\begin{equation}
\varepsilon^{\pm}_{1p}=\pm\frac{g_{e}\mu_{\rm B}B}{2}f_{1},\quad\,
\end{equation}
and the corresponding zeroth-order wave functions are
\begin{equation}
|\Psi^{\pm}_{1}\rangle=c^{\pm}_{1}|\Psi_{1\uparrow}\rangle+d^{\pm}_{1}|\Psi_{1\downarrow}\rangle,
\end{equation}
where
\begin{equation}
c^{\pm}_{1}=\frac{\cos(\theta-\varphi)\pm\,f_{1}}{\sqrt{2[f^{2}_{1}\pm\,f_{1}\cos(\theta-\varphi)]}},\quad\,d^{\pm}_{1}=
\frac{-i(1-2\eta^{2})e^{-\eta^{2}}\sin(\theta-\varphi)}{\sqrt{2[f^{2}_{1}\pm\,f_{1}\cos(\theta-\varphi)]}},
\end{equation}
with
\begin{equation}
f_{1}\equiv\,f_{1}(\eta,\theta-\varphi)=\sqrt{\cos^{2}(\theta-\varphi)+(1-2\eta^{2})^{2}e^{-2\eta^{2}}
\sin^{2}(\theta-\varphi)}.
\end{equation}
The corresponding first-order wave functions are calculated as
\begin{eqnarray}
|\Psi^{\pm}_{1p}\rangle&=&c^{\pm}_{1}|\Psi_{1\uparrow}\rangle+d^{\pm}_{1}|\Psi_{1\downarrow}\rangle+i\frac{\xi}{2}e^{-\eta^{2}}\sin(\theta-\varphi)\sum^{\infty}_{n\neq1}\frac{1}{n-1}\left[\frac{(\sqrt{2}i\eta)^{n-1}\sqrt{n!}}{(n-1)!}+\frac{(\sqrt{2}i\eta)^{n+1}}{\sqrt{n!}}\right]\nonumber\\
&&\times\left[(-1)^{n-1}c^{\pm}_{1}|\Psi_{n\downarrow}\rangle-d^{\pm}_{1}|\Psi_{n\uparrow}\rangle\right]\label{Eq_wavefunction}.
\end{eqnarray}

Up to now, we have analytically obtained the lowest four energy levels and the corresponding wave functions, $|\Psi^-_{0p}\rangle$, $|\Psi^+_{0p}\rangle$, $|\Psi^-_{1p}\rangle$, and $|\Psi^+_{1p}\rangle$, of a nanowire quantum dot by using degenerate perturbation theory. Assume that the electron in the quantum dot is initially in the ground state $|\Psi^{-}_{0}\rangle$. When an external a.c.~electric field $eEx\cos(2\pi\nu\,t)$ is applied to the nanowire quantum dot, it will induce coherent oscillations between the ground state and other excited states. We can calculate the transition matrix elements induced by the external a.c.~electric field between the $n=0$ sub-levels and the $n=1$ sub-levels:
\begin{eqnarray}
\langle\Psi^{-}_{0p}|x|\Psi^{+}_{0p}\rangle&=&ix_{0}\xi\eta\,e^{-\eta^{2}}\sin(\theta-\varphi),\nonumber\\
\langle\Psi^{-}_{0p}|x|\Psi^{-}_{1p}\rangle&=&\frac{\sqrt{2}}{2}x_{0}\left[(c^{-}_{0})^{*}c^{-}_{1}
+(d^{-}_{0})^{*}d^{-}_{1}\right],\nonumber\\
\langle\Psi^{-}_{0p}|x|\Psi^{+}_{1p}\rangle&=&\frac{\sqrt{2}}{2}x_{0}\left[(c^{-}_{0})^{*}c^{+}_{1}
+(d^{-}_{0})^{*}d^{+}_{1}\right]+i\frac{\sqrt{2}}{4}x_{0}\xi\eta^{2}e^{-\eta^{2}}\left[(c^{-}_{0})^{*}d^{+}_{1}
-(d^{-}_{0})^{*}c^{+}_{1}\right]\sin(\theta-\varphi),\label{Eq_transitions}\\
\langle\Psi^{+}_{0p}|x|\Psi^{-}_{1p}\rangle&=&\frac{\sqrt{2}}{2}x_{0}\left[(c^{+}_{0})^{*}c^{-}_{1}
+(d^{+}_{0})^{*}d^{-}_{1}\right]+i\frac{\sqrt{2}}{4}x_{0}\xi\eta^{2}e^{-\eta^{2}}\left[(c^{+}_{0})^{*}d^{-}_{1}
-(d^{+}_{0})^{*}c^{-}_{1}\right]\sin(\theta-\varphi),\nonumber\\
\langle\Psi^{+}_{0p}|x|\Psi^{+}_{1p}\rangle&=&\frac{\sqrt{2}}{2}x_{0}\left[(c^{+}_{0})^{*}c^{+}_{1}
+(d^{+}_{0})^{*}d^{+}_{1}\right].\nonumber
\end{eqnarray}
The transition selection rules are schematically shown in Fig.~\ref{Fig_SM}, where the transitions 1-5 successively correspond to the transition matrix elements given in Eq.~(\ref{Eq_transitions}). We note that in the absence of  spin-orbit coupling, i.e., $\alpha=0$, the Hamiltonian of a single nanowire quantum dot is $H=\frac{p^{2}}{2m_{e}}+\frac{1}{2}m_{e}\omega^{2}x^{2}+\frac{g_{e}\mu_{\rm B}B}{2}\sigma^{n}$. In this simple case, the spin of the electron is a good quantum number and the lowest four levels are just $\psi_{0}(x)|\uparrow_{n}\rangle$, $\psi_{0}(x)|\downarrow_{n}\rangle$, $\psi_{1}(x)|\uparrow_{n}\rangle$, and $\psi_{1}(x)|\downarrow_{n}\rangle$. Because of the spin degree of freedom, only transitions corresponding to those denoted by 2 and 5 in Fig.~\ref{Fig_SM} are allowed in the case of zero spin-orbit coupling. However, in the presence of spin-orbit coupling, i.e., $\alpha\neq0$, the electron spin is not a good quantum number any more, and both spin and orbital degrees of freedom of the electron are hybridized together [see Eq.~(7) of the main text and Eq.~(\ref{Eq_wavefunction}) herein]. As we show in Eq.~(\ref{Eq_transitions}), all the transitions 1-5 are allowed, and the transitions 1, 3, and 4 are only induced by the spin-orbit coupling.

The parameters $c^{\pm}_{n}$ and $d^{\pm}_{n}$ are the normalized coefficients. Thus, we have $|c^{\pm}_{n}|<1$ and $|d^{\pm}_{n}|<1$.  We can estimate that $|\langle\Psi^{-}_{0p}|x|\Psi^{-}_{1p}\rangle|<\sqrt{2}x_{0}$, $|\langle\Psi^{-}_{0p}|x|\Psi^{+}_{1p}\rangle|<\sqrt{2}x_{0}$, $|\langle\Psi^{+}_{0p}|x|\Psi^{-}_{1p}\rangle|<\sqrt{2}x_{0}$, and $|\langle\Psi^{+}_{0p}|x|\Psi^{+}_{1p}\rangle|<\sqrt{2}x_{0}$. The Zeeman splitting $g_{e}\mu_{\rm B}B$ is much less than the orbital splitting $\hbar\omega$, so the level spacings between the $n=0$ sub-levels and the $n=1$ sub-levels can be approximated as $\hbar\omega$. In implementing the EDSR, the frequency of the driving electric field is usually in resonance with the level spacing of the spin-orbit qubit, i.e., $h\nu=E_{\rm qu}$. However, this frequency is largely detuned with the level spacings between the $n=0$ sub-levels and the $n=1$ sub-levels. These detunings are approximated as $\hbar\omega-h\nu\approx\hbar\omega$ (see Fig.~\ref{Fig_SM}). Therefore, when the electric field for implementing the EDSR is weak, so that the condition
$\sqrt{2}eEx_{0}/\hbar\omega\ll 1$
is satisfied, the transitions between the $n=0$ sub-levels and the $n=1$ sub-levels are negligible due to the large frequency detuning~\cite{Scully_s}. For other sub-levels with $n=2,\cdots,\infty$, the frequency detunings are much larger, so the transitions between the spin-orbit Hilbert space and these states are even negligibly weaker. Therefore, when implementing the EDSR, the nanowire quantum dot can indeed be used as a spin-orbit qubit, i.e., a two-level quantum system. We take the InSb nanowire quantum dot as an example, where $\hbar\omega\approx7.5$ meV~\cite{Nadj2_s} and $m_{e}\approx0.014m_{0}$~\cite{Isaacson_s}, with $m_{0}$ being the free-electron mass. Corresponding to the condition $\sqrt{2}eEx_{0}/\hbar\omega\ll 1$, the amplitude of the applied electric field should satisfy $E\ll2\times10^{5}~V/$m. This gives an upper bound of the achievable Rabi frequency $\Omega_{\rm R}\sim 50$~GHz when choosing $\xi\sim 0.1$ and $\eta e^{-\eta^2}\sim 0.4$ (which is close to its optimal value). In order to realize coherent Rabi oscillations of the spin-orbit qubit, $1/\Omega_{\rm R}\sim 0.02$~ns should be much less than the relaxation time of the qubit. In experiments, this is easy to achieve because the quantum coherence of a spin-orbit qubit is even much better to satisfy this condition.

\begin{figure}
\includegraphics{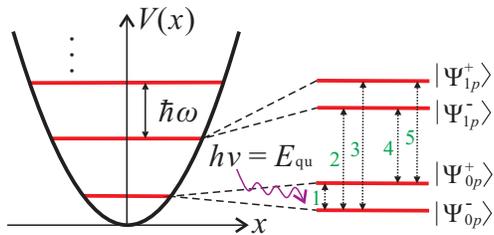}
\caption{\label{Fig_SM}Schematic diagram of the transition selection rules under an a.c. driving electric field among the lowest four energy levels in a single nanowire quantum dot. The transition 1 is shown in Eq.~(10) of the main text, and transitions 2-5 are shown in Eq.~(\ref{Eq_transitions}) herein. The frequency of the driving a.c. electric field is in resonance with the level spacing of the lowest two energy levels ($h\nu=E_{\mathrm{qu}}$).}
\end{figure}

Also, it should be noted that the Rabi frequency decreases after an optimal SOC strength $\eta_{\rm opt}=\sqrt{2}/2$ [see both Eq.~(12) and Fig.~2 in the main text]. This effect can be understood by analyzing the perturbation wave function that we have obtained. For the zeroth-order wave functions shown in Eq. (5) in the main text, there is no EDSR effect, because the zeroth-order wave functions only include the $n=0$ sub-level states. Thus, it does not contain the orbital degree of freedom of an electron in a quantum dot. Only the first-order perturbative wave functions given in Eq.~(7) in the main text contain both the orbital and spin degrees of freedom of an electron in a quantum dot, so it can respond to an external a.c.~electric field. As seen in Eq.~(7) in the main text, the exponential decrease $e^{-\eta^{2}}$ is included in this first-order perturbative wave functions.

For a single electron in a double quantum dot, Ref.~\onlinecite{Khomitsky_s} shows that the Rabi oscillations are slowed down when increasing the electric field. This is different from the effect we study here, because the slowing down of the Rabi oscillations in their case might result from the electron tunneling between the two dots. However, in our case, only a single quantum dot is involved and there is no electron tunneling, so the decrease of the Rabi frequency is purely due to the strong SOC.

\end{widetext}
\end{document}